\documentclass[10pt,twocolumn]{IEEEtran}
\usepackage{amsmath}
\usepackage{amsfonts}
\usepackage{epsfig}
\usepackage{amssymb}
\usepackage{cite}
\usepackage{hhline}
\usepackage{multirow}
\usepackage{framed}
\usepackage{xcolor}
\usepackage{lipsum}

\usepackage{color}
\usepackage{graphicx}
\usepackage{subfigure}

\begin{document}

\title{Energy Efficient Spectrum Sensing and \\ Handoff Strategies in Cognitive Radio Networks}

\author{Hossein Shokri-Ghadikolaei, Ioannis Glaropoulos, Viktoria Fodor, Carlo Fischione, and Konstantinos Dimou %
\thanks{H. Shokri-Ghadikolaei, I. Glaropoulos, V. Fodor, and C. Fischione are with KTH Royal Institute of Technology, Stockholm, Sweden (e-mail: {hshokri, ioannisg, vjfodor, carlofi}@kth.se).}
\thanks{K. Dimou is with Ericsson Research, Wireless Access Networks, Stockholm, Sweden (e-mail:  konstantinos.dimou@ericsson.com).}
\thanks{This work was supported by the FP7 EU project Hydrobionets.}}

\maketitle

\begin{abstract}
The limited spectrum resources and dramatic growth of high data rate communications have motivated opportunistic spectrum access using the promising concept of cognitive radio networks. Although this concept has emerged primarily to enhance spectrum utilization, the importance of energy consumption poses new challenges, because energy efficiency and communication performance can be at odds. In this paper, the existing approaches to energy efficiency spectrum sensing and handoff are classified. The tradeoff between energy consumption and throughput is established as function of the numerous design parameters of cognitive radio networks, both in the case of local and of cooperative spectrum sensing. 
It is argued that a number of important aspects still needs to be researched, such as fairness, dynamic behavior, reactive and proactive schemes for energy efficiency.
\end{abstract}

\section{Introduction}\label{Intro}
The great popularity of devices such as smartphones, tablets, and laptops all wirelessly connected to Internet as well as the recent development
of the Smart Grids paradigm has caused an exponential traffic increase over networks, which is exacerbating the overall global energy consumption.
Since wireless communications account for half the total energy consumption in ICT, energy efficiency is becoming a figure of merit for all wireless networks, which is now being researched in the area of green communications. Energy efficiency is also motivated by prolonging the battery lifetime of wireless devices, especially considering the increasing data rates. Therefore, there is a consensus in academia, industry, policy makers, and standardization bodies on the need of energy efficient wireless communications \cite{Feng2013Survey}.

In traditional wireless networks, energy efficiency considers the utilization of the available spectrum space, through appropriate protocol design and resource allocation. This is becoming impellent also in the emerging spectrum sharing techniques based on opportunistic access. Such techniques have been proposed to alleviate the spectrum shortage problem caused by the rapidly growing spectrum requirement of emerging wireless networks.
The cognitive radio network (CRN) is a key concept of opportunistic spectrum sharing, where secondary users (SUs), with cognitive capabilities, search for spectrum resources not used by the licensed ones, the primary users (PUs), and select their transmission and channel access parameters such that PU transmissions are not disturbed.

There has been a substantial effort in investigating CRNs with focus on throughput and interference management and optimization. Given that PUs' high data rates may cause high interferences to the secondary users, the energy consumption is even worsened in CRNs due to repeated attempts in channel sensing by the SU and spectrum handoffs. It fact, channel sensing is as much energy consuming as transmission. Therefore, in spectrum sharing CRNs, it is essential to address the joint optimization of energy resources spent for exploring the spectrum access opportunities and for transmission.

However, energy consumption cannot be considered as the only metric of figure in wireless networks in general and in CRNs in particular. Spectral efficiency is essential to achieve adequate quality of services. This is particularly important for spectrum handoff, which is the fundamental mechanism in a CRN, by which dynamic channel access and interference avoidance mechanisms allow the coexistence between primary and secondary systems. This motivates the investigation of the tradeoff between the spectrum utilization and the energy spent for transmission and reception in spectrum sensing and handoff.

In this paper, we investigate such a tradeoff and we propose a new figure of merit, the energy efficiency, defined as bit per Joule, namely the number of bits that can be transmitted per unit of energy. We review the spectrum sensing and spectrum handoff strategies proposed in the literature and evaluate the design parameters of CRNs affecting the energy efficiency metric.

\section{Fundamentals}\label{Fundamentals}
In this section, we summarize the essential aspects of CRNs, spectrum sensing, and handoff.
\subsection{Cognitive Radio Network}
\begin{figure}[t]
	\centering
		\includegraphics[width=8.5cm]{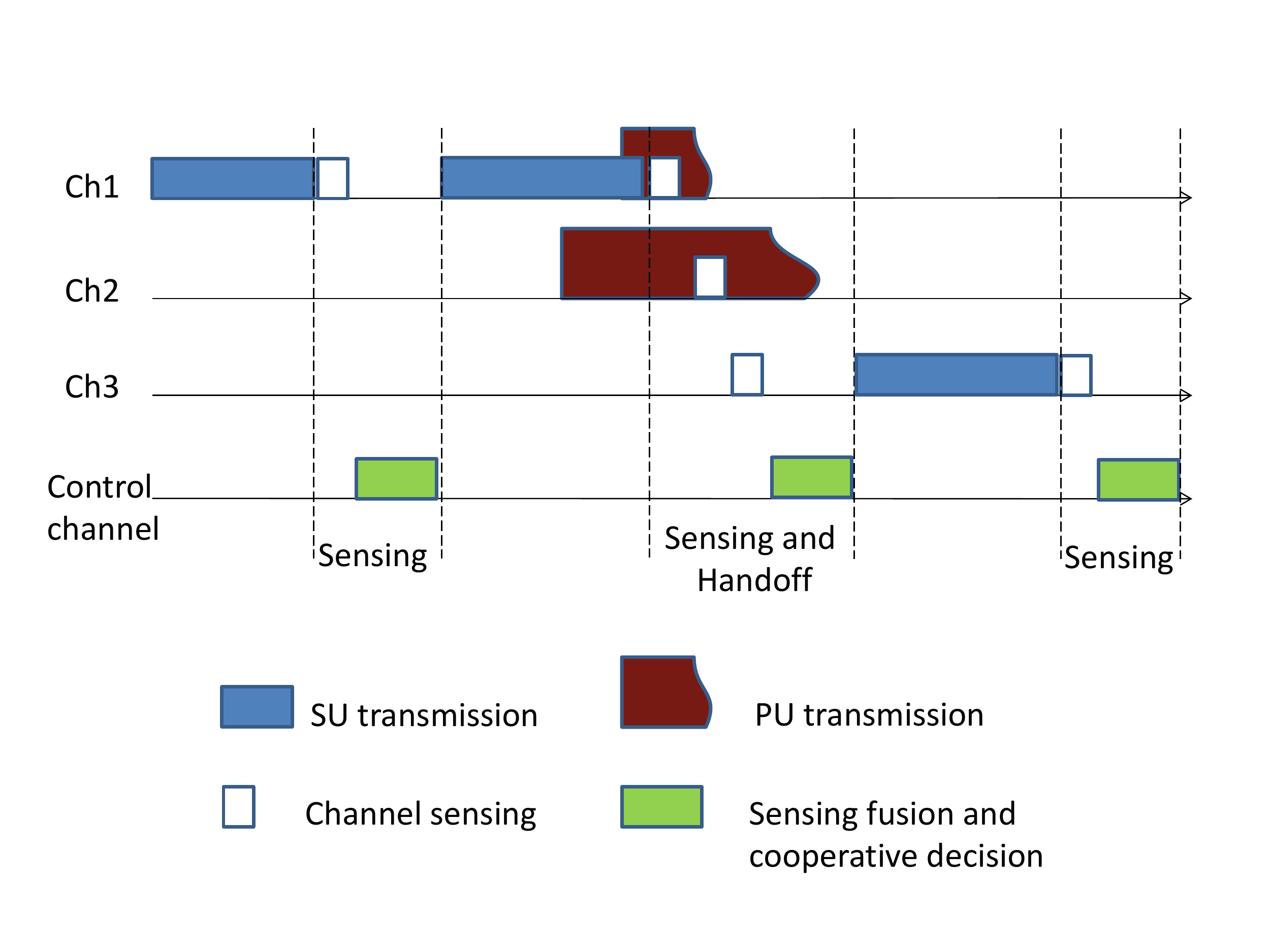}
	\caption{The SU interrupts transmission to evaluate the availability of the  channel using cooperative sensing. If the channel is busy, it starts a handoff procedure to find an idle channel. Spectrum sensing and handoff consumes both energy and time.}
	\label{fig:crn-sensing}
\end{figure}

Under opportunistic spectrum sharing, two or more networks share some part of the spectrum. The primary network owns the spectrum, and has performance guarantees. The secondary network(s) can access the spectrum if no significant degradation on the primary communication is caused. To find the opportunities of spectrum access, the secondary network learns the wireless environment and adapts to it. The learning is often based on sensing the spectrum, while the adaptation includes the tuning of the parameters of the protocols \cite{wang2011advances}.
As shown on Fig.~\ref{fig:crn-sensing}, to find the transmission opportunities appropriately and to protect the PUs from interference, the SUs need sensing the channels regularly using local or cooperative sensing, and start a spectrum handoff procedure, if the current channel is busy.

\subsection{Spectrum Sensing}
The most important parameters impacting  the performance of spectrum sensing are the time available to sense the transmission channels, the strength of the primary signals, and the a-priory information on the primary technology. The noise and the channel impairments such as shadowing and fading lead to decision errors, quantified in terms of false alarm and miss detection probabilities. A false alarm occurs when a free channel is mistakenly sensed busy, while a miss detection happens whenever an occupied channel is sensed free. To improve the sensing performance, cooperative sensing is introduced, where a group of SUs together decide about the availability of the channel.

\subsection{Multichannel Spectrum Sensing}
As typically there is more than one channel available for secondary access, spectrum sensing is divided into wideband and narrowband.
Under wideband spectrum sensing, an SU senses multiple channels simultaneously. Although this may allow short sensing duration, it requires complex hardware implementation, high power and computational complexity. Under narrowband sensing, only one channel can be sensed at a time, leading to easier implementation, lower power consumption, and less computational complexity. This is the reason why there has been great interest about such a sensing. By the narrowband sensing method, SU tries to find a free channel by sensing new channel sequentially according to a predetermined set of channels called sensing order. This has the potential drawback of a longer sensing duration.



\subsection{Spectrum Handoff Strategy}
The spectrum handoff strategy needs to answer the questions: When should an SU vacate the current channel? Should the SU wait on this channel or start finding another one? Which channels should be sensed and in what order?

The strategies can be categorized as reactive, proactive, or hybrid (a combination of the first two).  
Under reactive spectrum handoff, the SUs initiate searching among the channels to find transmission opportunities and pursue their unfinished transmissions, whenever handoff is required.
Although a larger delay becomes inevitable, reactive handoff leads to up-to-date channel status estimation.
Proactive schemes exploit the long term traffic statistics of the channels to establish a proper policy for future handoffs. 
As handoff can be carefully planned beforehand, very short handoff latency can be achieved. However, these schemes usually impose two radios on the SUs, one for transmission and another for channel observation. Also, as the sensing order is determined in advance, the predetermined target channels may not be longer available when handoff is triggered. Most of the proposed solutions follow a hybrid strategy combining the advantages of the two basic schemes. 

\subsection{Energy Efficiency}
Energy efficiency is generally defined as information bits transmitted per unit of energy, called "\emph{bit-per-Joule}" \cite{Feng2013Survey}.
This definition calls for a cross-layer optimization across all the protocol layers and networking functions, such as transmission and spectrum handoff in our case. The energy consumed in the secondary network consists of consumption for i) data transmission, ii)  spectrum sensing, iii) the communication protocol, including information exchange for cooperative spectrum sensing and for organizing the secondary transmissions.
Additionally, the circuit power consumed by the transmitter and the receiver, and the power consumed for tuning to the channel to be sensed can give a substantial part of the energy consumption \cite{PeiJSAC2011,WangVT12}.


By the Shannon's capacity formula, it is known that in dedicated spectrum, linearly increased transmission power leads to a logarithmic increase of the achievable transmission rate, and consequently the energy efficiency, as the ratio of the rate and the invested energy has an optimum value \cite{Feng2013Survey}. The power consumption and transmission rate tradeoff changes in cognitive radio networks due to the additional power consumption and also to the time needed for spectrum sensing and handoff.

In the following two sections, we analyze the energy efficiency of local and cooperative spectrum sensing, with focus on energy consumption. The key design parameters that affect the energy of these strategies are characterized.

\section{Local Spectrum Sensing and Handoff Strategies}\label{LocalSensing}
Local spectrum sensing can provide adequate sensing performance if the primary signals are sufficiently strong. Research on optimizing the spectrum handoff process under local sensing typically focuses on the SU and PU performance, in terms of maximizing the average throughput or similarly minimizing the secondary delivery time, for a single SU, or for a large secondary network. Below, we discuss how the mutual relationship between the key design parameters in the spectrum handoff process affect the energy.


\subsection{Channel Sensing Time}\label{SensingTimeSebSection}
The channel sensing time usually has a minimum value, which is required to achieve acceptable false alarm and miss detection probabilities. However, this minimum value does not lead to optimal energy efficiency. In fact, it affects the energy efficiency of the SU in a complex way. Increasing sensing time increases the energy consumption of that single sensing process, and decreases the time left for secondary transmission. However, as it also increases the probability of correctly detecting an idle channel, it leads to lower number of spectrum handoff, and consequently less energy consumption for sensing, and increased transmission time.

\begin{figure}[t]
	\centering
	\subfigure[]{
		\includegraphics[width=8.5cm]{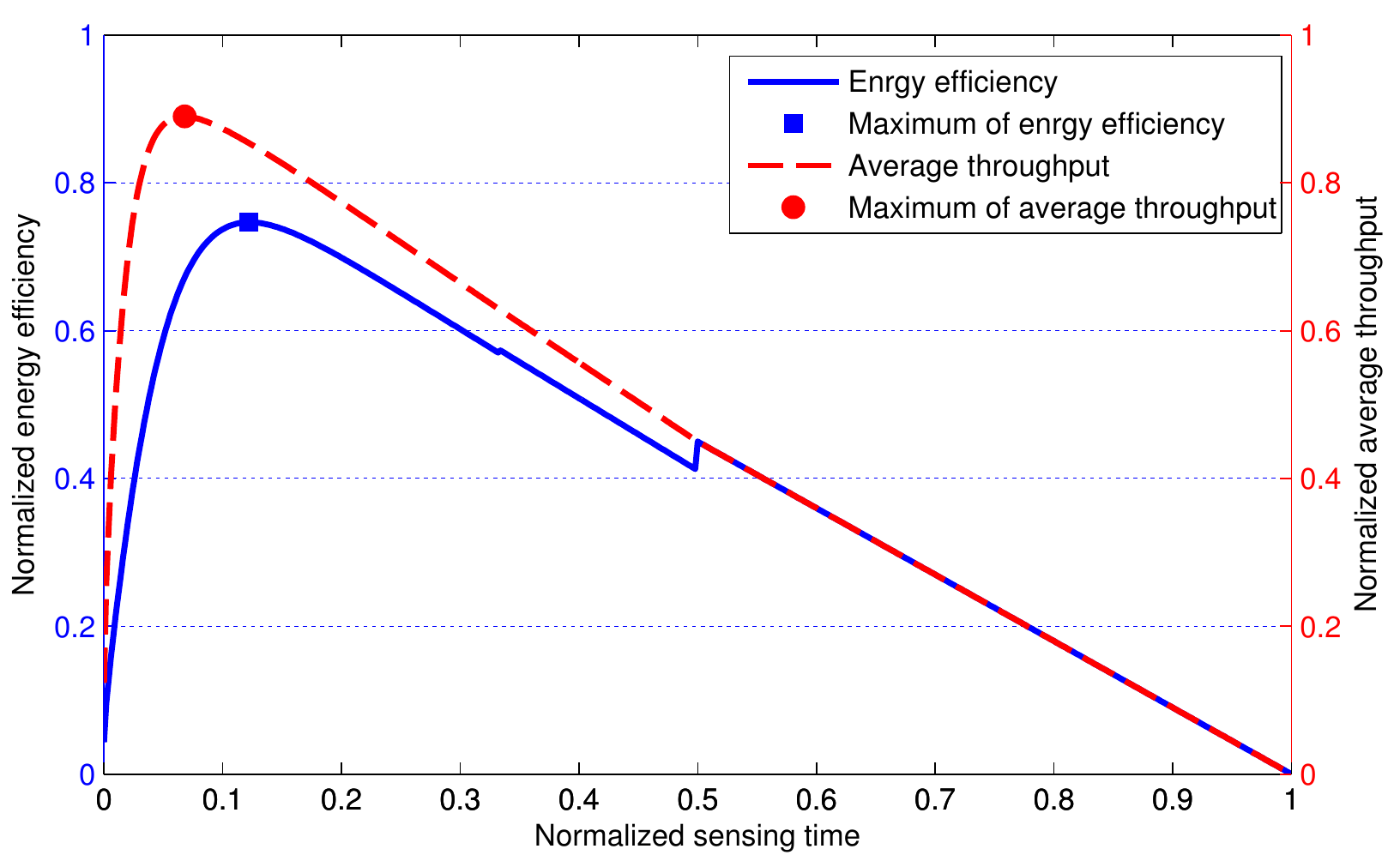}
		\label{subfig:SingleUserCase}
	}
	\subfigure[]{
		\includegraphics[width=8.5cm]{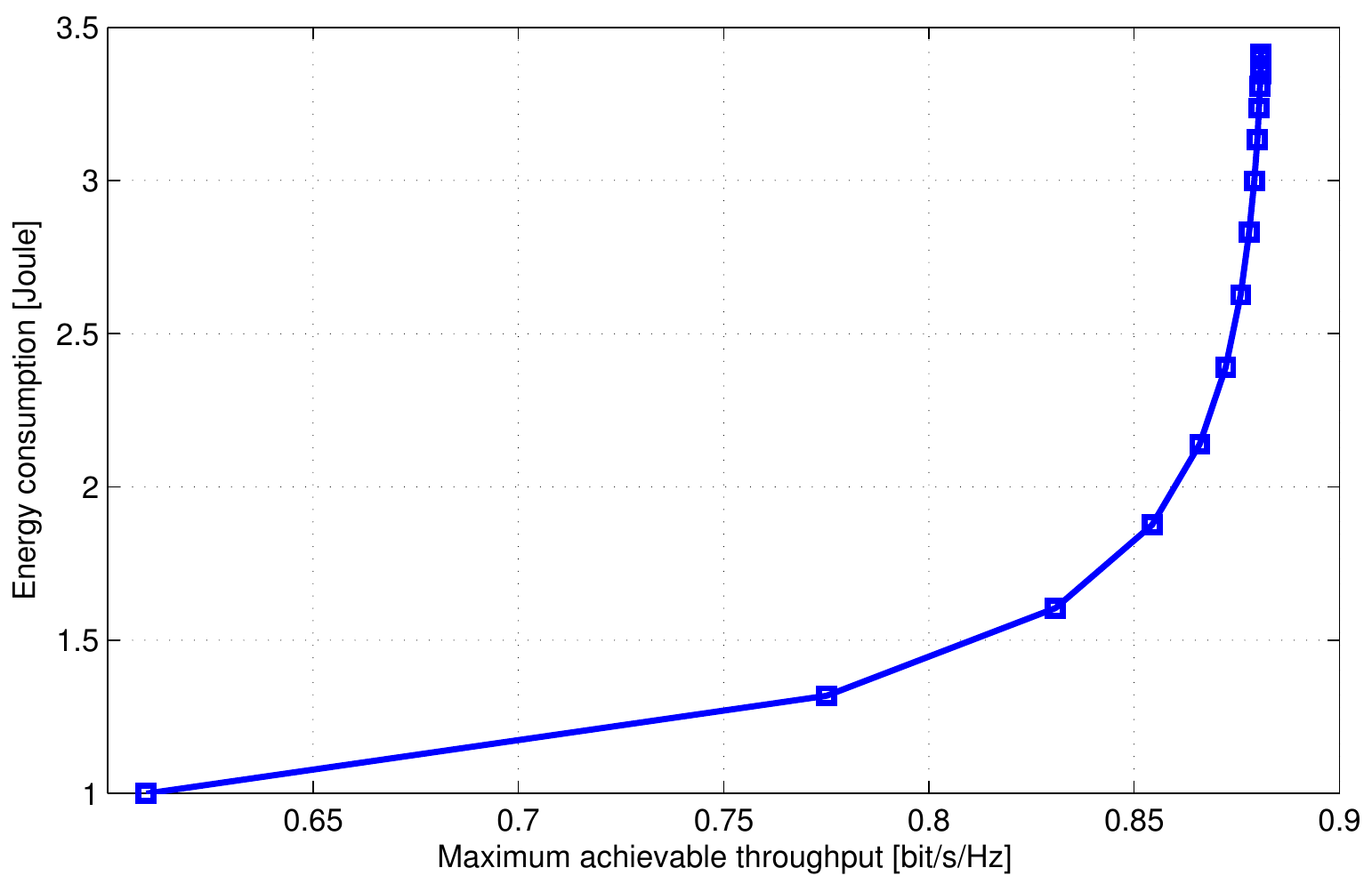}
		\label{subfig:EE_Thr}
	}	
	\caption{\subref{subfig:SingleUserCase}: Energy efficiency versus sensing time in a multi channel CRN with a single SU. The optimal sensing time considering energy efficiency is close to the throughput optimal one, but they are not the same.
   \subref{subfig:EE_Thr}: Throughput-energy tradeoff by increasing the maximum number of handoffs. The energy consumption needs to be increased significantly to utilize all the available spectrum resources. Based on \cite{ShokriIETC13}.}
	\label{fig:sensing-time}
\end{figure}

The tradeoff between sensing time, achievable throughput, and energy efficiency for a sequential spectrum sensing is evaluated in \cite{ShokriIETC13}.
As shown on Fig. \ref{subfig:SingleUserCase}, the energy efficiency first increases with the sensing time, due to a more accurate spectrum sensing, and reaches a maximum value. After the point, the energy efficiency falls, as the increased sensing performance can not compensate for the increased energy consumption and for the decreased time available for transmission. Comparing the optimal sensing time values for energy efficiency and for throughput maximization, we see that the optimal sensing time values for throughput or energy efficiency maximization are indeed different.

\subsection{Maximum Number of Handoffs}
The energy consumption of the spectrum handoff however, depends not only on the time used for sensing a single channel, but also on the number of channels that are sensed before an opportunity is discovered and the transmission starts.

Clearly, allowing the SU to discover more primary channels increases the chances of finding an empty channel and thus increases the throughput. However, as we see  on Fig. \ref{subfig:EE_Thr}, the energy consumption cost of this increase can be tremendous, once the system is close to the throughput limit. For instance, to increase the throughput above $0.85$, only $3\%$ transmission rate enhancement is achieved by $81\%$ more energy consumption, which devastates the energy efficiency. Therefore, the maximum number of handoffs needs to be carefully optimized based on the SU throughput requirements and available energy resources.

\subsection{Sensing Order}
With narrowband sensing, the SU sequentially senses the channels until an idle one is found. The order of channels to be sensed affects the average throughput and the average energy consumption.
As a result of an improper sensing order, an SU may sense several channels to find a transmission opportunity, and thereby may suffer from more energy consumption and less transmission time.
In \cite{PeiJSAC2011}, the energy efficient spectrum handoff for a CRN with one SU investigates a hybrid handoff strategy, wherein the SU learns the channel occupancy and transmission channel quality statistics, and defines the sensing order accordingly.
It is shown that optimizing only based on one of the above parameters can be highly sub-optimal, with a loss of energy efficiency up to $5-20\%$ for the considered scenario. Further, \cite{Nguyen2011Optimal} shows that significantly higher gain can be achieved by excluding from the sensing order the number with high occupancy probability or bad channel conditions.


\subsection{Sensing and Channel Access}
Finding an idle channel, however, does not guarantee successful transmission in secondary networks with several, uncoordinated users. Here, all SUs may sense the popular primary channels (like the ones with low load and good transmission quality), and then compete for accessing the same channel, while other channels might be idle too.

\begin{figure}[t]
	\centering
	\subfigure[]{
		\includegraphics[width=8.5cm]{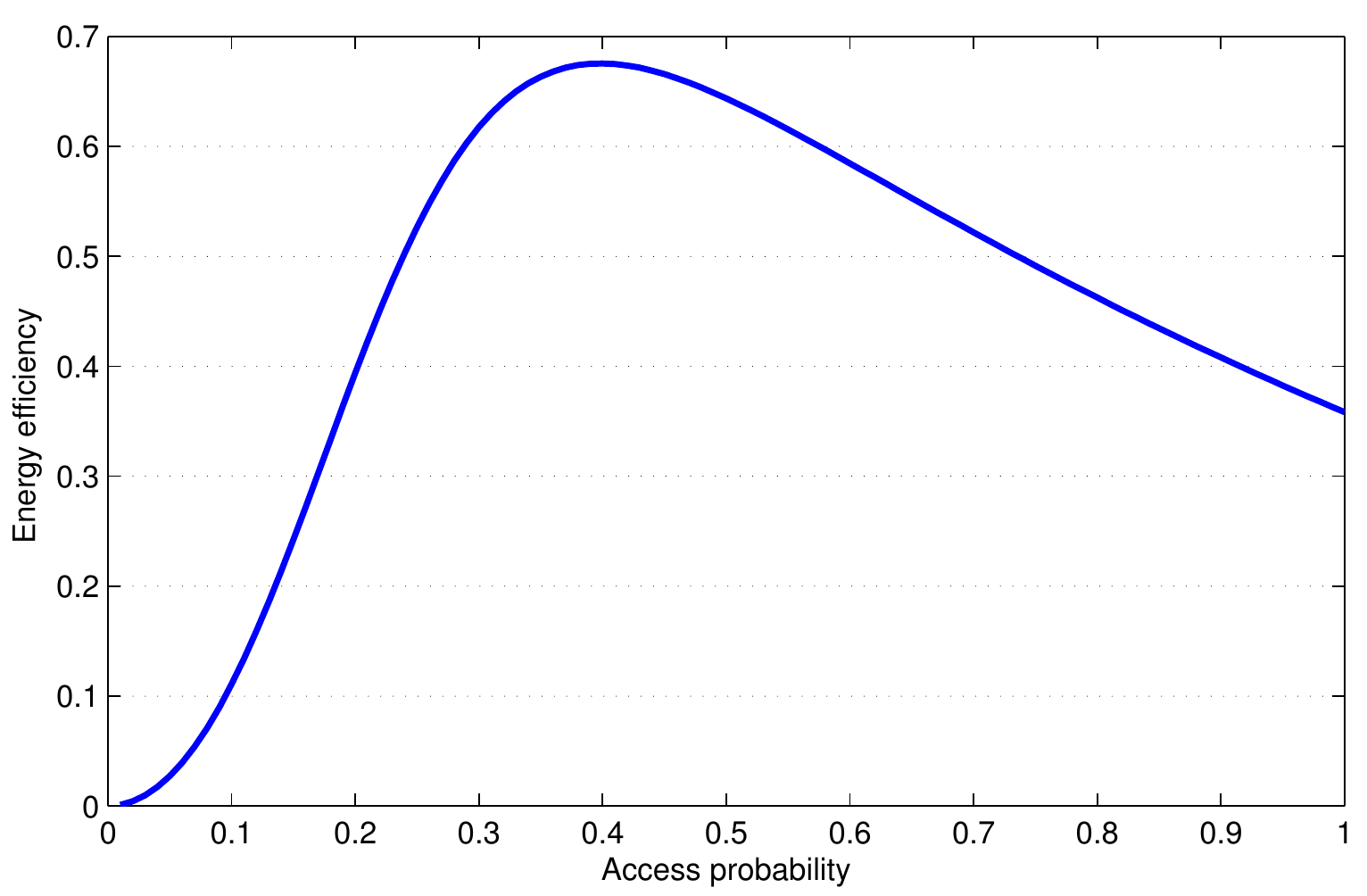}
		\label{subfig:EE_p}
	}
	\subfigure[]{
		\includegraphics[width=8.5cm]{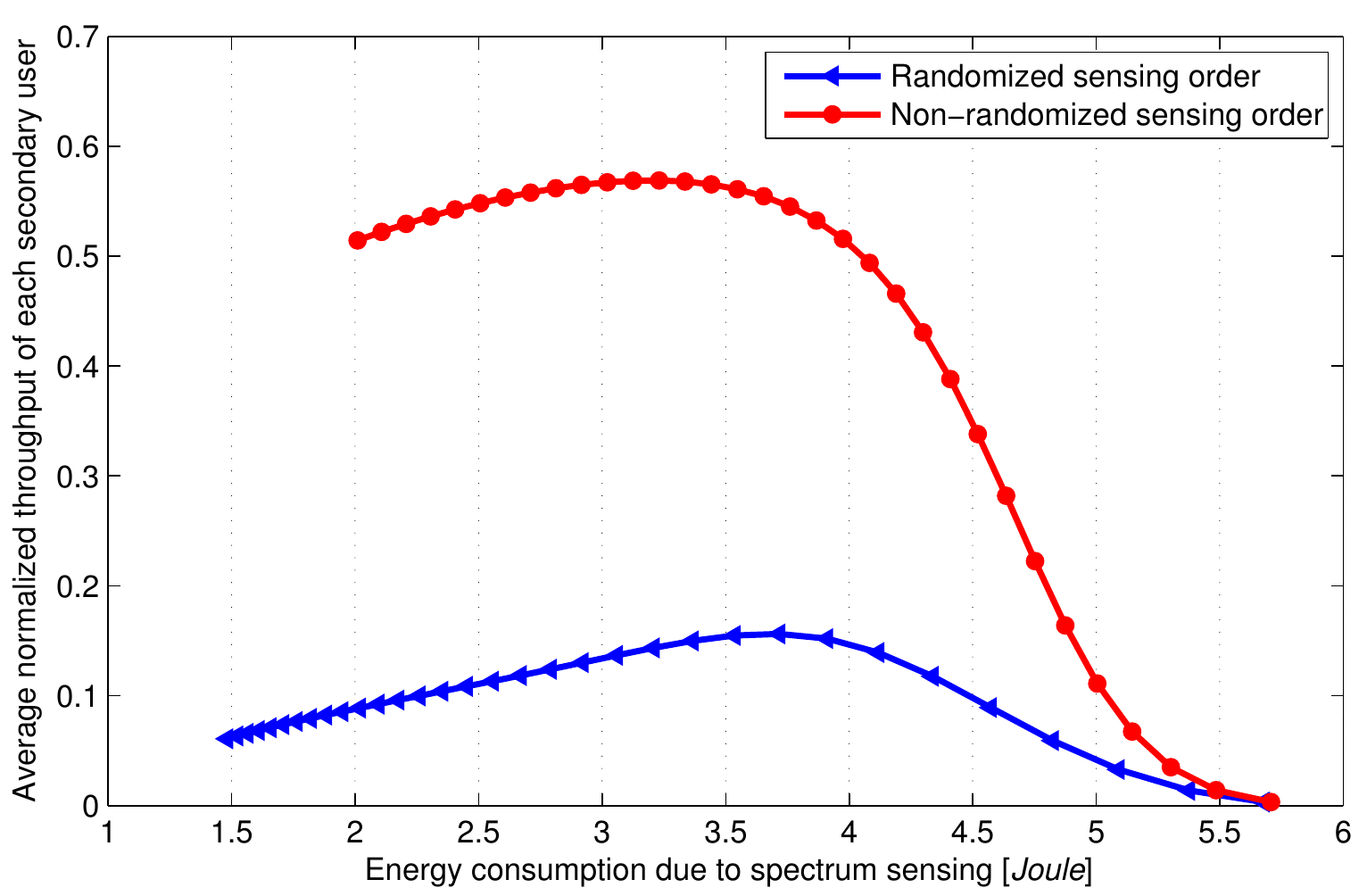}
		\label{subfig:thr_ho}
	}	
	\caption{\subref{subfig:EE_p}: Energy efficiency as a function of secondary access probability. Optimizing the access probability can improve energy efficiency significantly.
   \subref{subfig:EE_Thr}: Secondary throughput as a function of invested sensing energy. The joint access probability and randomization of the channel sensing order achieves significant throughput gain. Based on \cite{ShokriTWC13}.}
	\label{fig:sensing-access}
\end{figure}

To solve this problem, \cite{ShokriTWC13} suggests to couple sensing and channel access control. They introduce a randomized scheme, were the SUs sense and then access the channels with some access probability. As Fig. \ref{subfig:EE_p} shows, the access probability has a significant effect on the energy efficiency, due to the tradeoff between throughput enhancement at more intentions to access the channels and the consequent contention level increment. The optimum probability depends on the size of the secondary and primary networks, and should be tuned accordingly.
Further improvement can be achieved by randomizing the order of the channels to be sensed, as shown on Fig. \ref{subfig:thr_ho}. Thereby, the SUs are statistically distributed among all primary channels, and each SU can achieve substantially higher energy efficiency due to much lower contention level in the target primary channel.

Finally, \cite{song2012prospect} shows that in large CRNs, optimizing the access strategy jointly with sensing order, based also on the channel availability statistics, may be efficient. However, as high contention on popular channels needs to be avoided, the decision schemes become more complex.

%
%


\subsection{Waiting or Handoff}
Once a primary user returns to its channel, the SU may decide to wait until the channel becomes idle again, or invest some time and energy to start the handoff procedure and migrate to an idle channel.
As \cite{WangVT12} suggests, the decision should be based on the throughput and delay requirements of the SU. Unless the secondary quality requirements are very strict, optimizing the probability of waiting instead of migrating can decrease the energy consumption of the SU up to $20\%$. Clearly, the gain decreases as the throughput or delay requirements get strict, and the SU can not afford to simply wait for the new transmission opportunity in the current channel.

Given that a waiting SU needs to discover that the channel become idle again, \cite{zhang2013what} investigates how often the channel should be sensed.
More frequent sensing allows the SU to start to transmit with lower discovery delay and thus achieve higher throughput, at the same time, it requires higher energy consumption. \cite{zhang2013what} shows that sensing does not need to be periodic, and an adaptive sensing interval, based on some knowledge on the distribution of the PU busy time can halv the discovery delay, while keeping the same sensing energy budget as periodic sensing.

\section{Cooperative spectrum sensing}
In the case of strong primary signals, local sensing may be sufficient to ensure good performance. However, the cooperation of several spectrum sensing nodes, that is, secondary users in the area, is needed if the primary signal is weak, or if the radio propagation environment is harsh. There, the spatial diversity among the secondary users mitigates the effect of link impairments due to fading and shadowing phenomena, and the secondary users, together, can discover spectrum access opportunities.

The design factors discussed for local spectrum sensing can be further optimized in cooperative sensing scenarios, considering the wireless environment of the individual users. However, there are now additional open questions affecting the energy efficiency of the cooperating users, as we see in the following subsections.

\subsection{Sensing Resource Allocation}
\label{subsec_sensing_resource_allocation}
Under cooperative sensing the sensing resource is not
only the sensing time but also the set of secondary users cooperating to
discover a spectrum access opportunity. Considering a single channel, increasing
the number of cooperating users increases linearly the energy cost for detecting
channel availability. If several channels need to be sensed,
increasing the number of cooperating users means that each user needs to
sense more channels. This decreases the local sensing performance, or the time
available for secondary transmission, as discussed already for local sensing.
Therefore, the careful assignment of sensing duties to the existing secondary
users becomes a key design factor of CRNs.

Accordingly, \cite{saud2013} proposes an iterative solution to involve SUs in sensing, until the desired overall sensing performance in
terms of miss detection and false alarm probabilities are met for all
channels. Clearly, the gain of such strategy increases together with the number of SUs in the area, and therefore is important in dense secondary networks.

\begin{figure}[t]
	\centering
	\subfigure[]{
		\includegraphics[width=5.75cm]{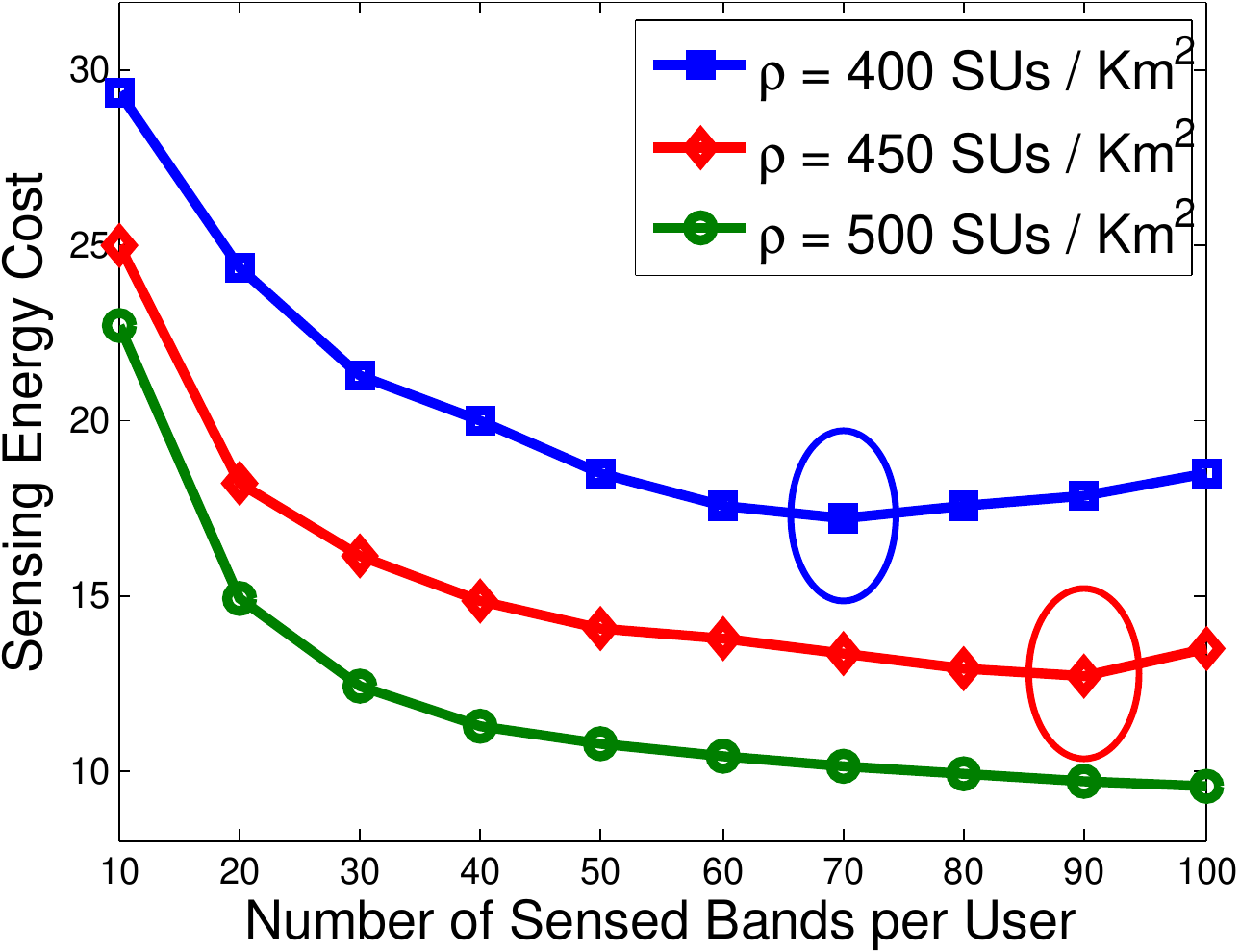}
		\label{subfig_energy_efficiency_vs_number_of_bands}
	}
	\subfigure[]{
		\includegraphics[width=6.15cm]{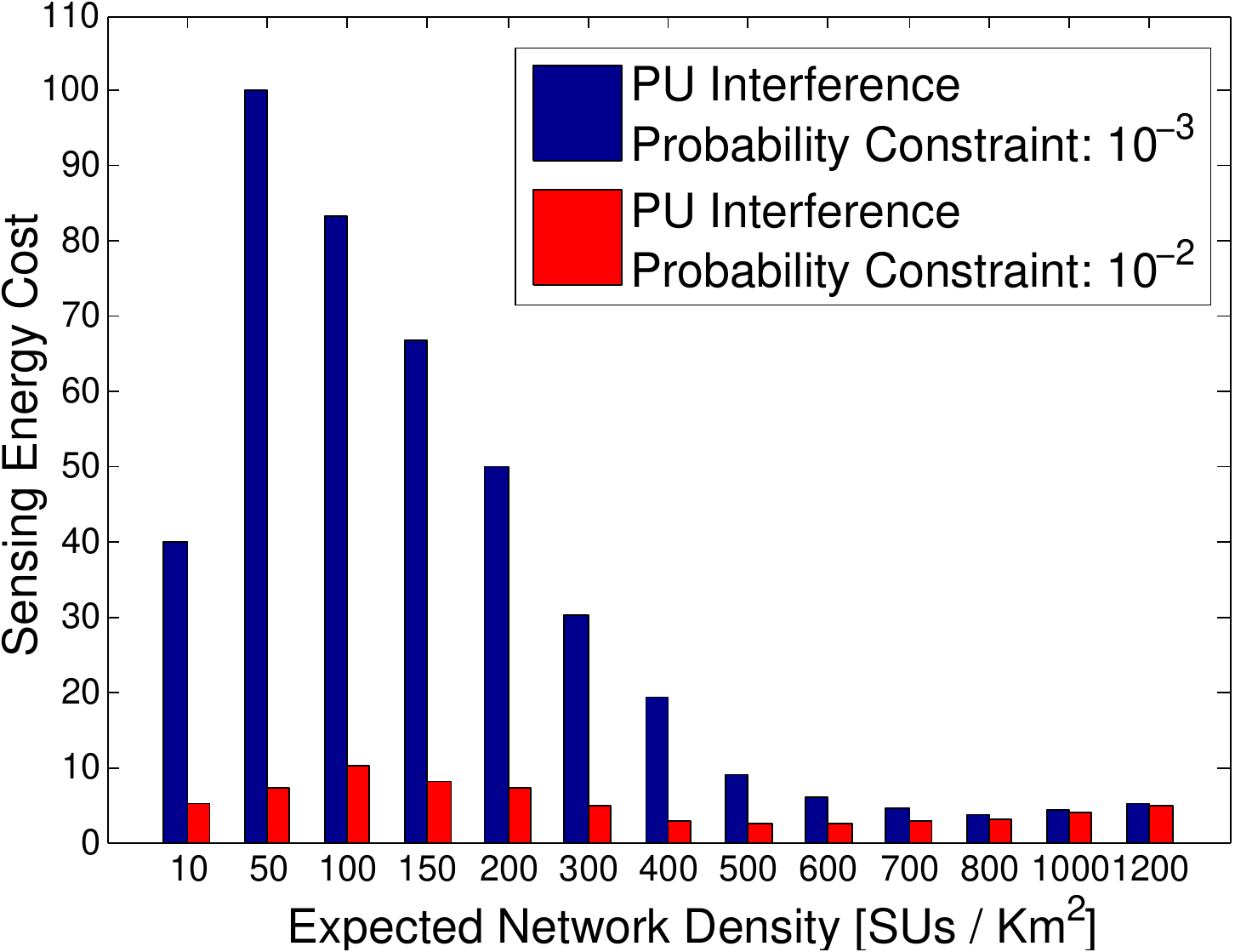}
		\label{subfig_energy_efficiency_vs_density}
	}	
	\caption{\subref{subfig_energy_efficiency_vs_number_of_bands}: The energy cost per
	unit of secondary use throughput is minimized when the number of sensed bands is optimal. The optimum depends on the user density.
	\subref{subfig_energy_efficiency_vs_density}: The energy cost is lowest
	at optimal cognitive network density, above which the sensing performance
	improvement does not compensate for the increased demand for cognitive
	capacity. Based on \cite{glaropoulos2009}. }
	\label{fig_glaropoulos}
\end{figure}

As discovering spectrum opportunities requires effort from a set of cooperating users, the SUs now need to decide how large part of the spectrum space, dedicated for secondary access, they want to utilize. On one side, they may increase the number of channels to
sense, so that there are more transmission opportunities to share. On the other side, this requires more sensing efforts from each SU. This shows that there has to be an optimal spectrum space to be sensed both for maximising per user throughput or energy efficiency.  This optimal value is evaluated in \cite{glaropoulos2009}, where it is shown that the density of the secondary network is an important design factor
(Fig.~\ref{subfig_energy_efficiency_vs_number_of_bands}) to minimize the energy
cost of each individual user. Such a cost is the sensing energy invested by the user
to gain a unit of transmission opportunity. Moreover, the energy cost, even if
minimized, strongly depends not only on the primary network quality
requirements, but also on the density of the secondary network. Networks with
moderate density are worst off, where the cooperative sensing performance is
moderate, but the gained access opportunities has to be shared by a large
set of nodes (Fig.~\ref{subfig_energy_efficiency_vs_density}).

\subsection{Sensing Coordination}
\label{subsec_sensing_coordination_scheduling}

\begin{figure*}[t]
	\centering
	\subfigure[]{
		\includegraphics[width=5.75cm]{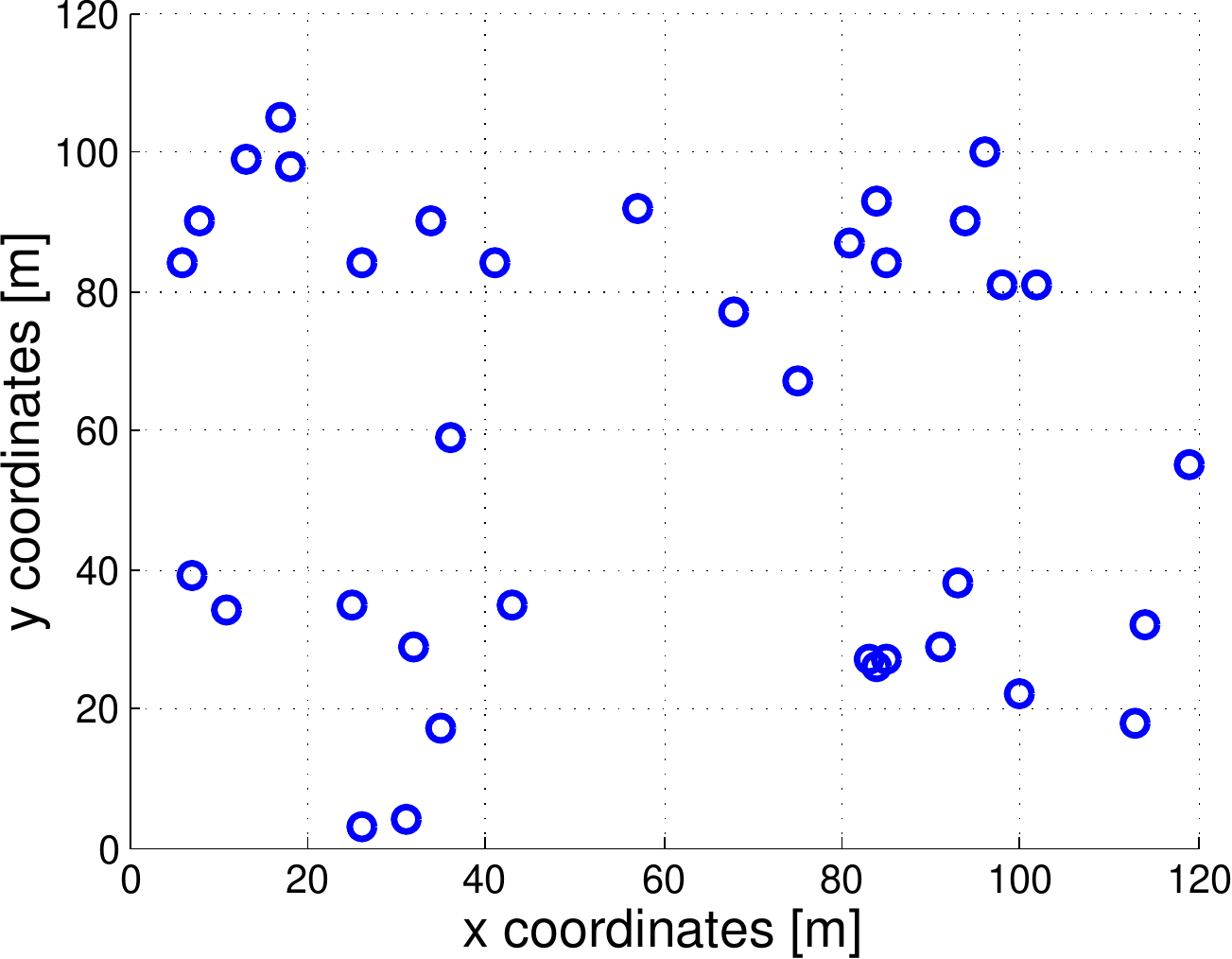}
		\label{subfig_topology_1}
	}
	\hspace{0.5cm}
	\subfigure[]{
		\includegraphics[width=5.75cm]{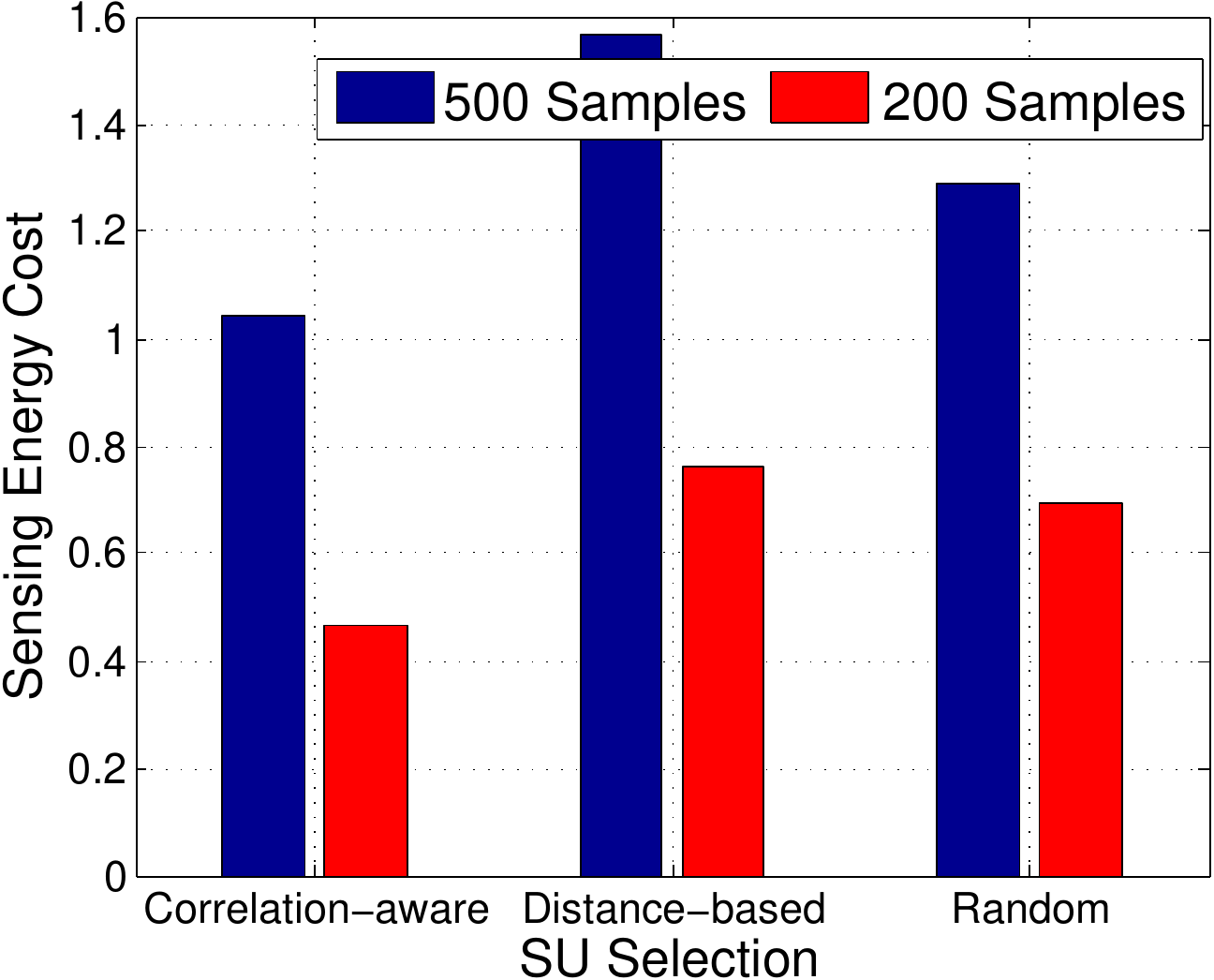}
		\label{subfig_energy_cost_vs_selection_1}
	}	
	\subfigure[]{
		\includegraphics[width=5.75cm]{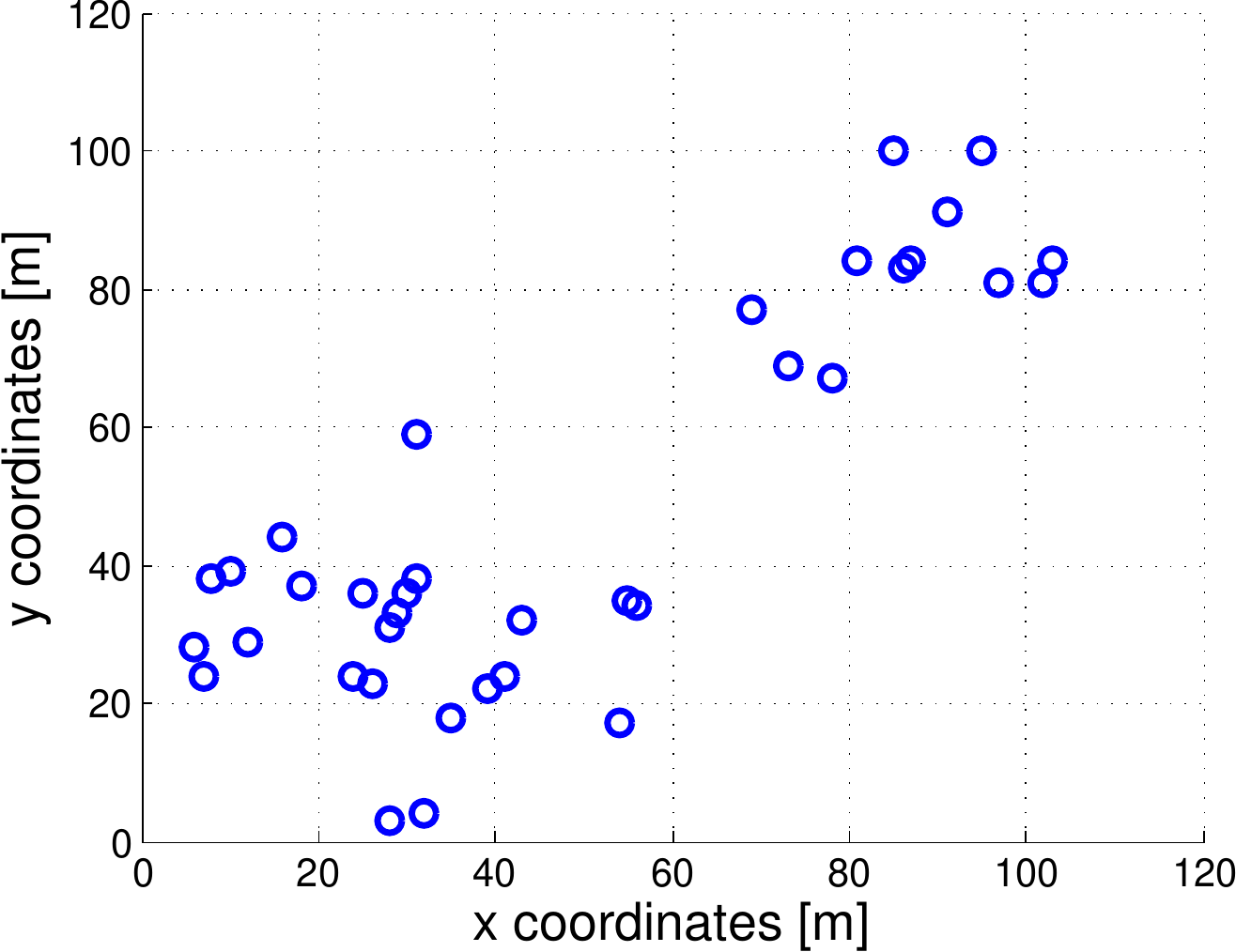}
		\label{subfig_topology_2}
	}
	\hspace{0.5cm}
	\subfigure[]{
		\includegraphics[width=5.75cm]{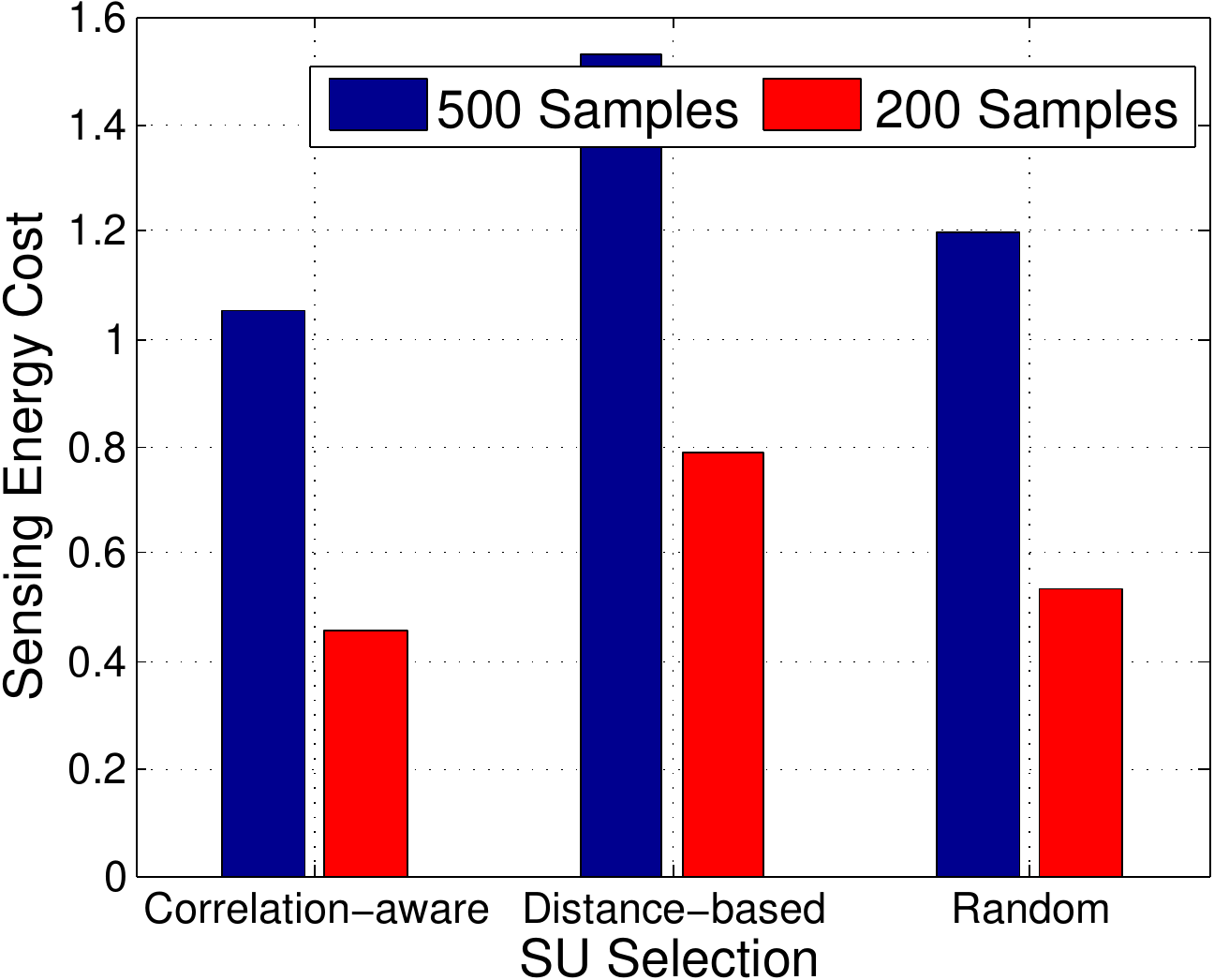}
		\label{subfig_energy_cost_vs_selection_2}
	}
	\caption{The energy cost per
	unit of secondary user throughput decreases when the correlation between SU
	measurements is taken into consideration in the user selection process. The
	improvement compared to random selection is smaller in
	\subref{subfig_energy_cost_vs_selection_2} since the nodes are located in
	disjoint geographical areas \subref{subfig_topology_2}. The higher sensing
	throughput, as a result of the increase int the sensing time per channel (500
	samples), does not compensate for the linear increase in the sensing energy
	overhead. Based on \cite{cacciapuoti2012}. }
	\label{fig_correlation_aware_selection}
\end{figure*}

Since we have shown that the number of users participating in the cooperative sensing needs to be carefully selected, the remaining issue is what should be considered when selecting which particular users cooperate. Given that the main reason for cooperative sensing is to mitigate fading and shadowing, \cite{cacciapuoti2012} suggests that users experiencing uncorrelated link attenuation should be selected. As shown in Fig.~\ref{fig_correlation_aware_selection}, the efficiency of this correlation aware policy depends on the spatial distribution of the SUs. It can decrease the number of users required to sense the primary channel, and, consequently, the sensing energy consumption by more than 50\%, without affecting, or even improving, the sensing performance, hence, the achieved throughput by the secondary system.

Another factor to consider when selecting the users to cooperate is the cost of collecting the results and making the cooperative decision.

\begin{figure}[t]
	\centering
	\subfigure[low SNR]{
		\includegraphics[width=6.0cm]{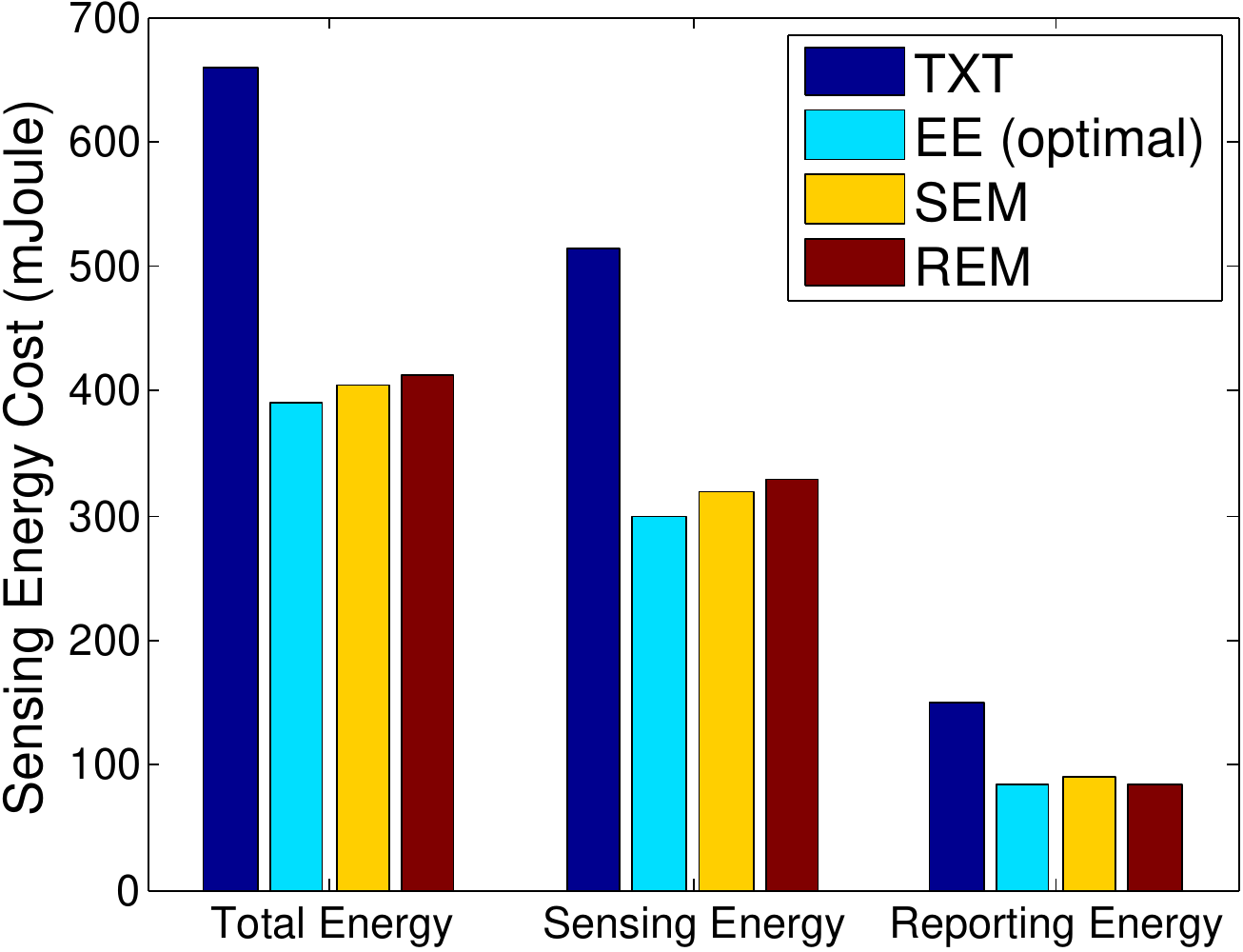}
		\label{subfig_energy_efficiency_vs_selection_low_snr}
	}
	\subfigure[high SNR]{
		\includegraphics[width=6.0cm]{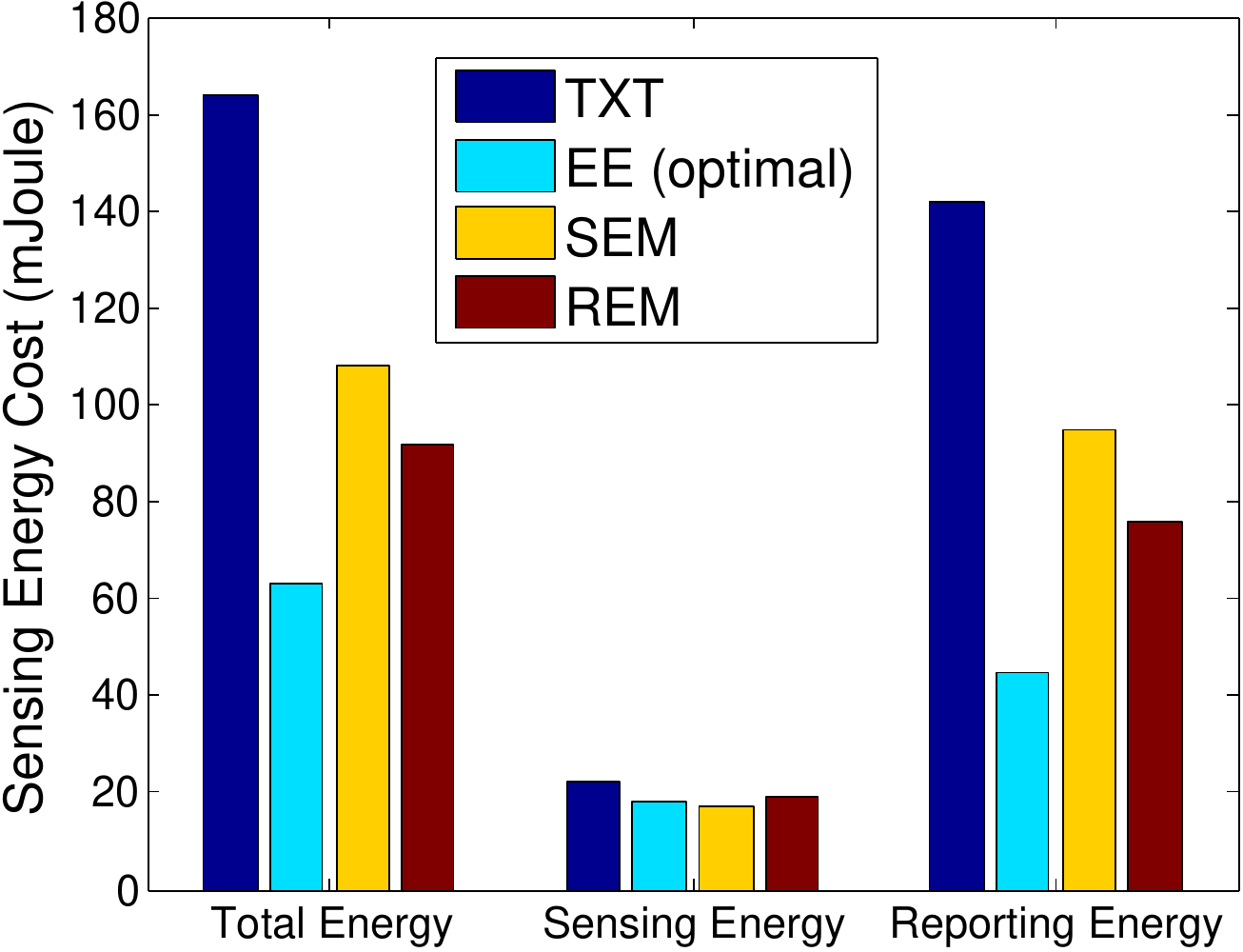}
		\label{subfig_energy_efficiency_vs_selection_high_snr}
	}	
	\caption{Cooperative sensing energy cost for the optimal and sub-optimal
	user selection schemes \cite{salim2013}. In case of low SNR regime
	\subref{subfig_energy_efficiency_vs_selection_high_snr} the sensing energy
	dominates the total energy consumption due to an increased required sensing
	time, while it drops significantly for high SNR
	\subref{subfig_energy_efficiency_vs_density}. The optimal (EE) SU selection
	scheme outperforms the heuristic solutions (SEM, where the optimization
	is performed considering only the sensing energy cost and REM, when sensing
	assignment is prioritized over users with already assigned sensing duties), and
	the TXT scheme where the sensing time is fixed, regardless of the received SNR
	at the SUs.
	The performance difference is more significant in the case of high SNR on the
	primary-secondary user links, when sensing itself costs little energy. Based on \cite{salim2013}.}
	\label{fig_energy_cost_vs_selection}
\end{figure}

Under cooperative sensing, the reporting of the local sensing results may require a significant amount of energy and even time, particularly if the links used for reporting requires a high transmission power or needs to be transmitted on a multihop path.
Therefore, \cite{salim2013} compares different ways to select the cooperating secondary users, considering the local sensing performance, the sensing result transmission cost, or both, with an objective to minimize the total required sensing energy cost for maintaining an overall sensing quality. As shown on Fig.~\ref{fig_energy_cost_vs_selection}, the gain of joint optimization  is significant, if the sensing itself does not require a significant amount of energy, due to good channel conditions or high primary transmission power.

\subsection{Sensing Report Forwarding}
\label{subsec_sensing_report_forwarding}

As reporting the sensing results may have significant cost, \cite{maleki2011} suggests that the SUs, even if included in the cooperative sensing, should
choose not report the sensing ersult, if it might have little impact on the cooperative decision, while it would raise the overall reporting cost. They show, that if the primary channel utilization statistics are a-priori known to the SUs, then the individual users can have a good estimate on the usefulness of their sensing result. In this case censored reporting can drastically reduce the total sensing energy overhead by up to 40\%, while maintaining the desired sensing performance.

%

\subsection{Decision Combining}
\label{subsec_decision_combining}

In addition to introducing a significant overhead to the overall energy cost of collaborative sensing, the reporting of the individual sensing results imposes a threat to the cooperative sensing performance due to the inherent lack of reliability of the wireless links used for reporting. Consequently, the cooperative decision has to mitigate the impact of the unreliability of the local sensing measurements and the errors introduced by the unreliable links.

To this end, \cite{chaudhari12} demonstrates that the quality of cooperative decision based on the individual decisions of the users (so called hard decision combining) can degrade by up to 60\% if the reporting links are unreliable. Instead, using cooperative decision based on quantized raw sensing results (that is, soft decision combining) the overall sensing performance can be maintained at a relatively high level. The granularity of the reported sensing results need to be tuned carefully to tradeoff the energy and delay of reporting and the throughput gain due to correct spectrum decisions.

\section{Discussions}
Spectrum handoff is a fundamental mechanism in CRNs, by which dynamic channel access and interference avoidance mechanisms allows for a smooth coexistence between primary and secondary systems. The optimization of the spectrum handoff aims primarily at increasing the efficiency of secondary access to underutilized primary channels, while maintaining a moderate energy overhead, so to enhance the lifetime of energy-constrained cognitive networks.

The energy efficiency of spectrum sensing depends heavily on the  maximum number and the order of the primary channels sensed by a secondary user, the frequency of the spectrum sensing, and the selection of the per-channel sensing time. As shown in Section \ref{SensingTimeSebSection}, the sensing time must be carefully adjusted to not let the energy resources and transmission opportunities be wasted for a marginally higher accuracy of spectrum sensing. In multi-user scenarios, the selection of the number and the order of the channels to be sensed becomes even more important. 

The problem of energy efficiency becomes more complicated in harsh wireless environments, where cooperative spectrum sensing is inevitable. As the energy cost of spectrum sensing is increased, maintaining the energy efficiency factor at moderate levels requires a careful optimization of the sensing resource allocation and coordination schemes. In Section \ref{subsec_sensing_resource_allocation}, we show that the allocation of sensing tasks to secondary users with relatively good individual sensing and uncorrelated channel conditions substantially reduces the energy consumption. A careful reporting and combining of the individual sensing results increases the overall sensing efficiency. This leads to higher energy efficiency, when taking into account the individual cost of reporting the sensing measurements, thus, allocating sensing to users with low-cost reporting links.

\section{Open Issues}
Although several interesting investigations exists in the area of energy efficient spectrum sensing and handoffs, many important issues are still open.
\begin{itemize}
  \item \textbf{Fairness in energy efficiency}: In heterogeneous secondary networks, it is essential to optimize the design parameters so to guarantee a fair allocation of sensing duties to the SU, each having diverse throughput requirements and/or energy resource constraints. Thus, the focus of cooperative spectrum sensing should be shifted from globally maximizing sensing performance or optimizing resource allocation to energy efficiency of CRNs, while maintaining individual performance demands.

  \item \textbf{Dynamic energy efficient design}: The current researches on energy efficiency are mainly based on static knowledge of the primary network model. Considering the dynamic spatio-temporal changes of the primary network, it is desirable to integrate dynamic learning techniques into an energy efficient spectrum sensing and handoff mechanisms. The energy efficiency of the adaptive sensing and handoff schemes should be analyzed taking into account the complexity, the overhead, but also the potential impact of the learning techniques in the overall efficiency of the CRN.

  \item \textbf{Proactive or reactive}: Although there are several proactive and reactive spectrum sensing and handoff proposals, a general comparison in terms of energy efficiency is missed. Based on the available information from the primary and secondary networks constraints such as the QoS level that should be guaranteed for the primary network, it would be very interesting to have a general framework for comparing different types of spectrum sensing and handoff and answering the following question: When does proactive spectrum handoff outperform reactive one in term of energy efficiency? Clearly, the answer might be different for single, centralized, and distributed cognitive radio networks, and also it might vary for local and cooperative spectrum sensing.
\end{itemize}

\bibliographystyle{IEEEtran}
\bibliography{IEEEabrv,bibfile}
\end{document}